\documentclass[fleqn,10pt]{wlscirep}
\usepackage[utf8]{inputenc}
\usepackage[T1]{fontenc}
\usepackage{tikz}
\usepackage{subfig}
\newcommand\mm[1]{#1}
\newcommand\mmm[1]{#1}
\newcommand\sk[1]{#1}

\title{Cauliflower shapes of bacterial clusters in the off-lattice Eden model for bacterial growth in a Petri dish \mm{with an agar layer}}

\author[1]{Szymon Kaczmarczyk}
\author[1]{Filip Koza}
\author[1]{Damian Śnieżek}
\author[1,*]{Maciej Matyka}

\affil[1]{
Faculty of Physics and Astronomy\\
Institute~of~Theoretical~Physics\\ 
		University~of~Wroc\l aw\\
	
pl.~M.~Borna~9, 50-204 Wroc\l aw, Poland}

\affil[*]{maciej.matyka@uwr.edu.pl}


\begin{abstract}
We developed the off-lattice Eden model to simulate the growth of 
bacterial colonies in the three-dimensional geometry of a Petri dish. In contrast to its two-dimensional counterpart, our model takes a three-dimensional set of possible growth directions and employs additional constraints on growth, which are limited by access to the nutrient layer. We rigorously tested the basic off-lattice Eden implementation against literature data for a planar cluster. We then extended it to three-dimensional growth. Our model successfully demonstrated the non-trivial dependency of the cluster morphology, non-monotonous dependency of the cluster density, and power law of the thickness of the boundary layer of clusters as a function of the nutrient layer height. Moreover, we revealed the fractal nature of all the clusters by investigating their fractal dimensions. Our density results allowed us to estimate the basic transport properties, namely the permeability and tortuosity of the bacterial colonies. 
\end{abstract}
\begin{document}

\flushbottom
\maketitle

\thispagestyle{empty}

\section{Introduction}

Despite their microscopic size, bacteria play fundamental roles in many biological, ecological, and industrial processes. Their ability to grow and adapt rapidly makes them the subject of numerous scientific studies, mainly because of their crucial importance in applications such as food production \cite{bacteriafood}, biotechnology \cite{aneja2007experiments}, and medicine \cite{Chiang2020}. Understanding pathogenic bacterial colony growth mechanisms is crucial for developing more effective infection control strategies. In the food industry, it is essential to optimize the fermentation process to achieve higher product quality and safety. In biotechnology, knowledge of the bacterial colonization processes is crucial to support the development of new methods for producing bioactive compounds, enzymes, and antibiotics \cite{Singh2017, Medema2011}. However, bacterial colony growth is not completely random, but follows the strict laws of biology and physics \cite{Shapiro95}. Experimental investigation of bacterial colony growth is expensive and time-consuming, and often requires special experimental protocols \cite{Wimpenny79}.
Computer modeling of bacterial colony growth predicts the behavior of microorganisms under different conditions without the need to prepare measurements\cite{Ben1994,Li1995}. A continuous reaction-diffusion equation was utilized to model the spatiotemporal growth of bacterial colonies \cite{Grimson}. Particle kinetic growth models, such as the Eden model \cite{Eden61}, Diffusion-Limited Aggregate DLA \cite{Witten81}, and kinetic growth processes \cite{AUSLOOS19952185} mimic the dynamics of growth, cell-cell interactions, and the influence of environmental factors on bacterial colony growth by simplifying the underlying physics and biology into a few basic processes of deposition and random cell motion. Computer modeling is often restricted to two-dimensional cases, where bacterial colony growth is known to occur in a three-dimensional space. 
A Petri dish is an experimental instrument that allows the growth and study of bacterial colony structures in the laboratory. It is a standard closed glass container that allows observation and control of bacterial growth under laboratory conditions.

In this study, we developed a new, simple model for the growth of bacterial cultures in a nutrient layer deposited in a Petri dish. For this purpose, we use an off-lattice Eden model based on \cite{Wang1995}. First, a detailed validation and two-dimensional case study were performed. We then generalize the model to three dimensions by extending the allowed growth direction set. Our contribution concerns the implementation of the model in the geometry of a Petri dish reduced to growth in a restriction by the surfaces of the nutrient-rich layer thickness. Using the new model, we studied the cluster growth dynamics across the area for different nutrient layer thicknesses. We show how the density of the clusters changes as a function of thickness and analyze their morphology in terms of the boundary layer thickness and fractal dimension.

\section{The model}

\begin{figure}
\centering
\includegraphics[width=0.4\columnwidth]{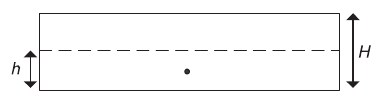}
\includegraphics[width=0.4\columnwidth]{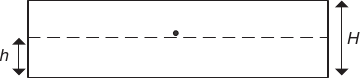}
\caption{\mm{Diagram of a Petri dish model with the growth starting in the middle (left) and on top (right) of the agar layer.} The dish has height $H$ in which the height of the nutrient-rich \mm{(agar)} layer \mm{$h=H/2$}  limits the thickness of the growing cluster. The bacteria is seeded in the middle \mm{(left)} \mm{or} on top \mm{(right)} of the food layer from which the cluster starts to develop\label{fig1-petri}.}
\end{figure}

The bacterial colony growth process can be divided into different phases \cite{monod}. Phase succession is closely linked to the presence of food in the growth environment. For growth in a Petri dish, the nutrient layer may be filled with agar medium in solid, liquid, or gel form. In our model, we assumed that new bacteria can grow only in areas rich in nutrients. The height of the food layer above and below the embryo limits its growth (see Fig.~\ref{fig1-petri}).  
The growth in the model started from a single cell deposited in the middle of the nutrient-rich layer of height $h$. To model the growth of a three-dimensional bacterial colony, we developed a meshless Eden-C model\cite{Wang1995}. We generalized the model to the third dimension while keeping the main procedure the same as in the original. \mmm{In our approach, we neglect the interaction between the cells and the agar. This means we assume either that the agar is the only medium the cluster can growth in or it acts as a nutrient source for cluster growing on top of it (see Fig.~\ref{fig1-petri}). }
\begin{figure}
\centering
\includegraphics[width=0.24\columnwidth]{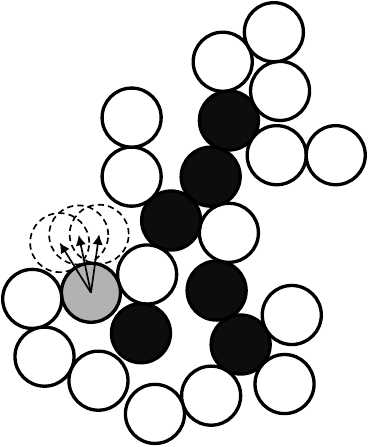}

\caption{The scheme of the cluster growth algorithm. Dead cells (black) have too little space around them to allow growth. They are surrounded by live boundary cells (white). The randomly selected live cell (marked in gray here) can grow in one of the allowed directions (schematically presented as vectors) in one of the possible locations for a new cell (dashed lines).
The positions of the cells in this scheme were taken from a snapshot of the actual simulation performed using the presented growth model.
\label{fig2-petri}}
\end{figure}

First, we define a living cell as one that has food available in free space. This means that there is some non-zero probability that a new cell will grow and attach to it during the growth process. To identify possible growth spots around the cell, we generated a discrete array of all possible growth directions (see Fig.~\ref{fig2-petri}). Subsequently, by random selection, the possible growth directions were chosen and tested against possible growth. The complete algorithm of our model based on the basic off-lattice Eden-C \cite{Wang1995} reads

\begin{enumerate}
\item[1.]  One living cell was selected (see Fig.~\ref{fig2-petri}) randomly.

\item[2.]  One direction was pulled out (randomly) from the array of possible directions. Then, we test whether new cells can appear.

\item[3.] If the cell cannot grow in the selected spot (the space is already occupied by live or dead cells or the selected spot is beyond the nutrient zone), $\rightarrow$ returns to step 2 (do this until all possible directions are checked); otherwise, $\rightarrow$ adds a new live cell and returns to step 1.

\item[5.] If all directions are pulled out and no more cells can grow $\rightarrow$, transform the cell into a "died" state and return to step 1. 
\end{enumerate}

The algorithm continues as long as there are living cells in a cluster. 
Our implementation uses a Mersenne Twister random number generator. For the best performance, the cell diameters were fixed at 1. 
Preliminary calculations were performed using a computer equipped with a single processor. The final calculations for averaging over independent colonies were performed on the 40-CPU computing cluster using independent, parallel runs of single instances of the program for changed parameters. 
Based on the observation of the variability of the results of the thickness of the boundary layer of the computed clusters with the number of discretized directions, we used 360 possible angles in all computations. On one hand, this guarantees a relatively fast algorithm and, on the other hand, adequate accuracy (and lack of variability) of the results obtained.

%
%
%

%
%
The complete C++ implementation of the above algorithm is available as Open Source \cite{code}.

\section{Results and Discussion}

\begin{figure}[!h]
\centering
\includegraphics[width=0.3\textwidth]{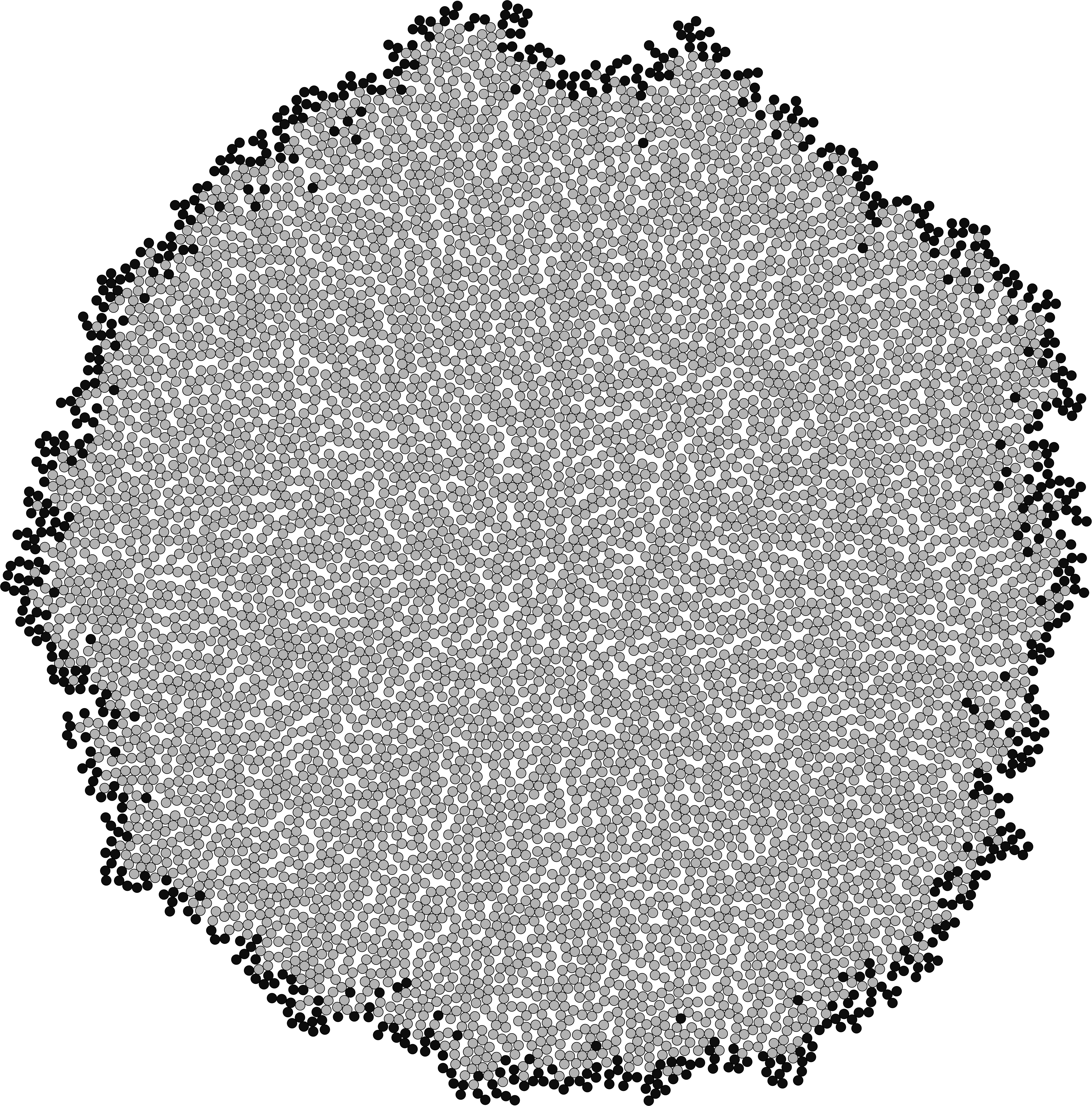}
\includegraphics[width=0.3\textwidth]{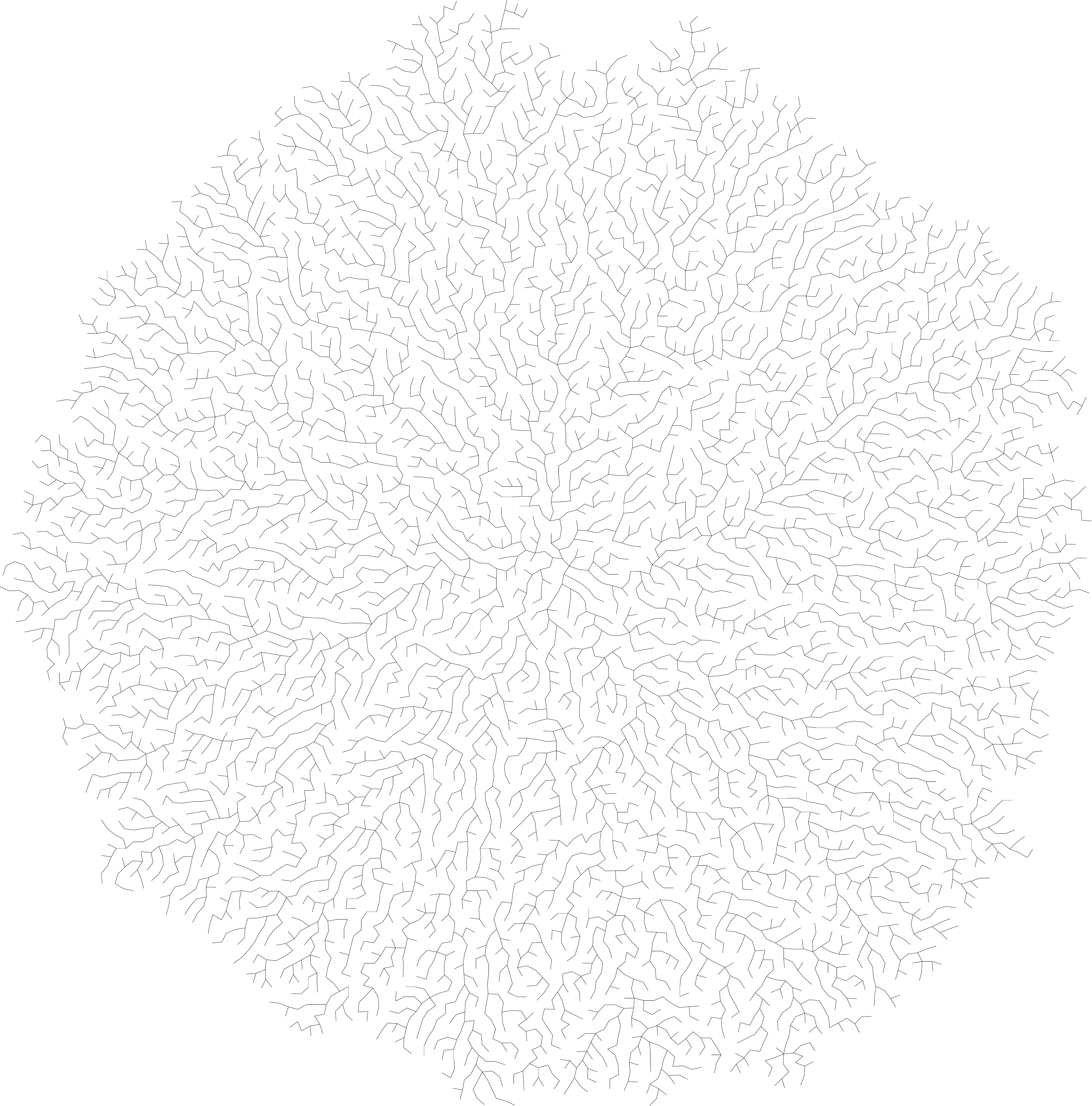}
\caption{The arrangement of cells embedded in the two-dimensional model for the single cluster realization (left). The growth network corresponds to the connections of successive growing cells to the actual cluster (right). \mmm{The total number of cells in the cluster is N=7599.}\label{clusters}}
\end{figure}

We first measured the classical cluster in the Eden model in two dimensions to verify our implementation, based on the original article \cite{Wang1995}. The intermediate objective was to investigate the two-dimensional structure, which would later become an extreme case of the three-dimensional Petri dish model.
An example cluster resulting from the algorithm above, along with the cluster growth network, is shown in Fig.~\ref{clusters}.

\subsection{Cluster density}

To measure the cluster density, we first define the dimensionless radius $R$. To obtain this, we averaged the distance between all living cells and the center of mass of the cluster. 
\begin{equation} 
R = \frac{1}{N_L}\sum_{i=0}^{N_L}  \eta(i)\frac{1}{r}\mid \vec{x}_i - \vec{x}_c \mid, 
\label{meanRadius}  
\end{equation} 
where $r$ is the single-cell radius, and the center of mass $\vec{x}_c$ is equal to 
\begin{equation} 
\vec{x}_c = \frac{1}{N_a}\sum_{i=0}^{N_a} \vec{x}_i .
\end{equation} 
Function $\eta(i)$ contains information regarding the state of the $i$th cell:
\begin{equation} 
\eta(i) =  
\begin{cases} 
    1 \quad\text{if}\quad \text{cell}_i \quad\text{is alive}\\ 
    0 \quad\text{if}\quad \text{cell}_i \quad\text{is dead} 
\end{cases}. 
\end{equation} 
We used the ratio of the number of cells to their squared radii to measure the cluster density. We know that the area of $N_a$ cells of radius $r$ is proportional to the location of the entire cluster of radii $R$. 
\begin{equation} 
N_a\pi r^{2} \propto \pi \left(Rr\right)^{2},   \end{equation} 
where radius $R$ of the entire cluster is expressed in units of r.
We simplify the following: 
\begin{equation} 
N_a\propto R^{2},
\label{eq::Ncell}
\end{equation} 
Furthermore, using the known formulas for the cell area and circle, we can write 
\begin{equation} 
\rho_s = \frac{N_a\pi r^{2}}{\pi \left( Rr \right) ^{2}}, 
\end{equation} 
which is equal to 
\begin{equation} 
\rho_s = \frac{N_a}{R^{2}}. 
\end{equation} 
Hence, for a homogeneous cluster, the density $\rho_s$ is equal to the slope of the straight line fitted to the measurement points in the graph $N_a(R^2)$ (see Fig.~\ref{fig2d-verification}). We obtained $\rho_s=0.65$ from our implementation, which perfectly corresponds with the result obtained there\cite{Wang1995}.

\subsection{Density of living cells layer}

\begin{figure}[!h]
\centering
\includegraphics[width=0.33\textwidth]{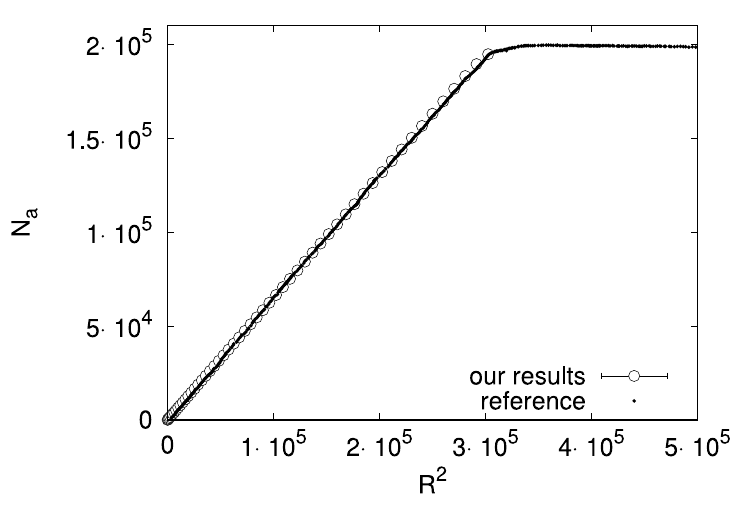}
\includegraphics[width=0.33\textwidth]{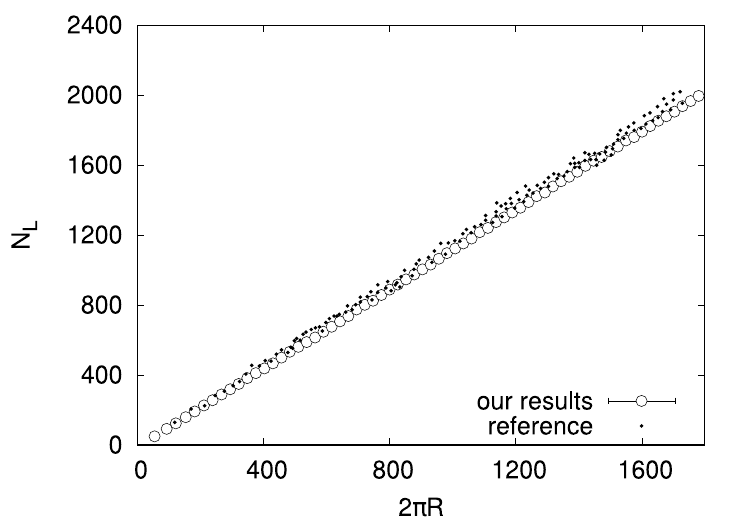}
\includegraphics[width=0.33\textwidth]{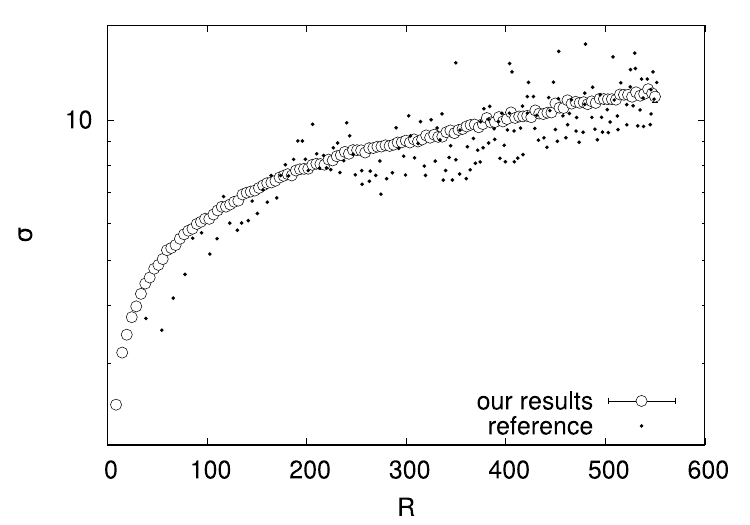}
\caption{Left: dependency of the number of cells in the function of the cluster diameter squared. The measurement points are the average results of 100 simulations, and the error bars are smaller than those of the measurement point symbols. Middle: Dependence of the number of living cells on the cluster boundary on its circumference. Right: Standard deviation (thickness) of the boundary layer in the cluster on the average radius.
Our results (open circles) were compared to reference\cite{Wang1995} (dots).
\label{fig2d-verification}}
\end{figure}

The linear density of a layer of living cells $d$ is defined as the number of living cells $N_L$ per unit length of the perimeter of a cluster: 
\begin{equation} 
d = r\frac{N_L}{2\pi R},  
\label{eq:lineardensity}
\end{equation} 
Plots of the average linear density of a layer of living cells are shown in Fig.~\ref{fig2d-verification}. We obtained the linear density $d=1.169$, which differs from\cite{Wang1995} (there, $d = 1.15$). These differences may occur because of the different number of space discretization points in the direction selection algorithm for embedding, which was not specified in the original study \cite{Wang1995}.

\subsection{Boundary layer thickness }

To determine the thickness of the boundary layer of a cluster composed of live cells, we used the standard deviation $\sigma$ of the distance from the center of the live cells \cite{Wang1995}. According to the formula:

\begin{equation} 
    \sigma\left( R \right) 
    =\frac{1}{r}\sqrt{\frac{1}{N_L}\sum_{i=0}^{N_L}\eta(i)\left( R - \left| 
    \vec{x}_i-\vec{x}_c \right|  \right)^{2} }, 
\end{equation}

The standard deviation was measured according to the current cluster size, which is expressed as $R$ according to formula \ref{meanRadius}. We assume that the relation describing $\sigma$ is of the following form\cite{kardar}: 
\begin{equation} 
    \sigma = \phi rR^{\gamma}.
    \label{eq:thickness}
\end{equation}  
By fitting the above relation to the simulation data, we obtain
$\phi\approx 1.169\pm 0.014$ and $\gamma\approx 0.360\pm 0.002$,  
that differs from the original study ($\gamma_{org}=0.396$). However, our results agree better with the values predicted for two-dimensional cluster interfaces $\gamma=\frac{1}{3}$ \cite{kardar}.

\subsection{Growth in a Petri dish}

Next, we ran a generalized model of the growth of a three-dimensional bacterial cluster in a Petri dish. \sk{The main difference between the two- and three-dimensional models is the possible directions, which are uniformly distributed around the circle in the two-dimensional case. In contrast, in a three-dimensional model, they may be obtained in several ways, for example consecutively selecting two random angles on the sphere or by selecting angles from regular distribution on the sphere. Depend on the choice it may even lead to a non-uniform distribution of directions on the sphere, which may be related to nutrient gradients near the surface of the bacterial colony that drive the growth process\cite{Martinez2022}.}

\sk{
To develop 3D model from 2D, we choose one of the possible ways to pick up grow directions:
\begin{itemize}
\item A random, independent choice of two angles (model \textbf{A})
\item A random choice one of the previously defined directions, from which each was a combination of two angles same as angles in model \textbf{A} (model \textbf{B})
\item A random selection of one of the predefined directions in such a way that they are uniformly distributed over the surface of the sphere (model \textbf{C})
\end{itemize}}
\begin{figure}[!h]
	\centering
	\includegraphics[width=1\columnwidth]{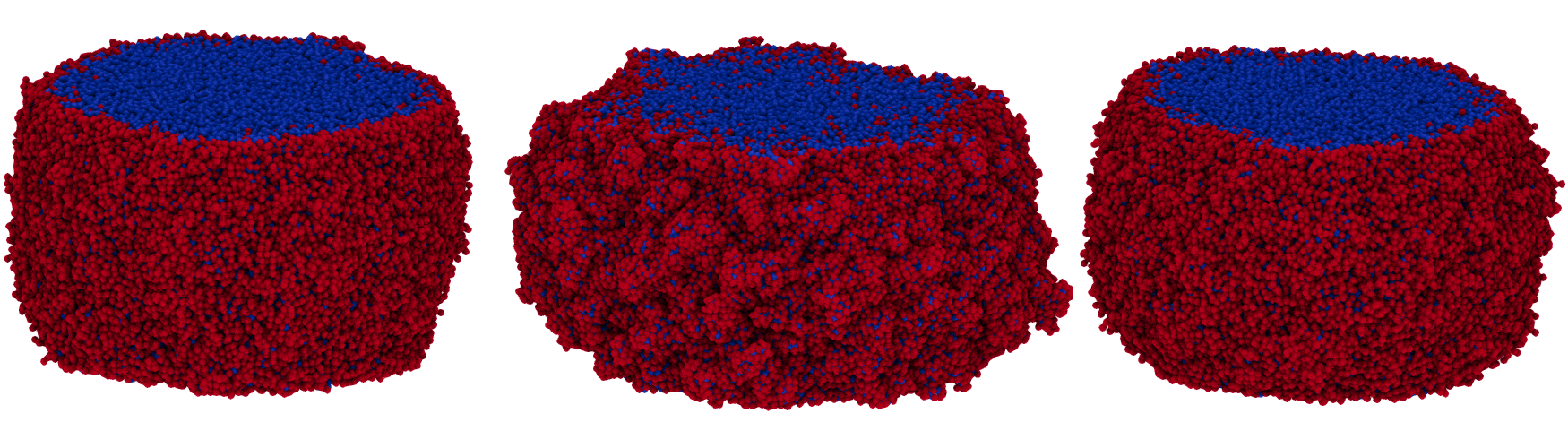}
	\caption{\sk{The final cluster configurations obtained with three distinct ways of chosing the direction of growth: model \textbf{A}, model \textbf{B}, model \textbf{C} (as described in text).}\mmm{The total number of cells in each cluster reads (from left to right): N=103808, 116150, 104649.}}
    \label{fig:modelComp}
\end{figure}
\sk{Based on the shapes of generated sample clusters (see Fig.~\ref{fig:modelComp}), we chose model \textbf{B} for further research as the most similar to natural cell's clusters\cite{Huergo2012, Colter2001, Kaufman2001} with special attention given to cauliflower-like shapes reported recently\cite{Martinez2022}. Additionally, we check what happens if we change the starting point of cluster growing from the middle of the agar layer to the top of the layer.}
\begin{figure}[!h]
    \centering
    \includegraphics[width=0.99\columnwidth]{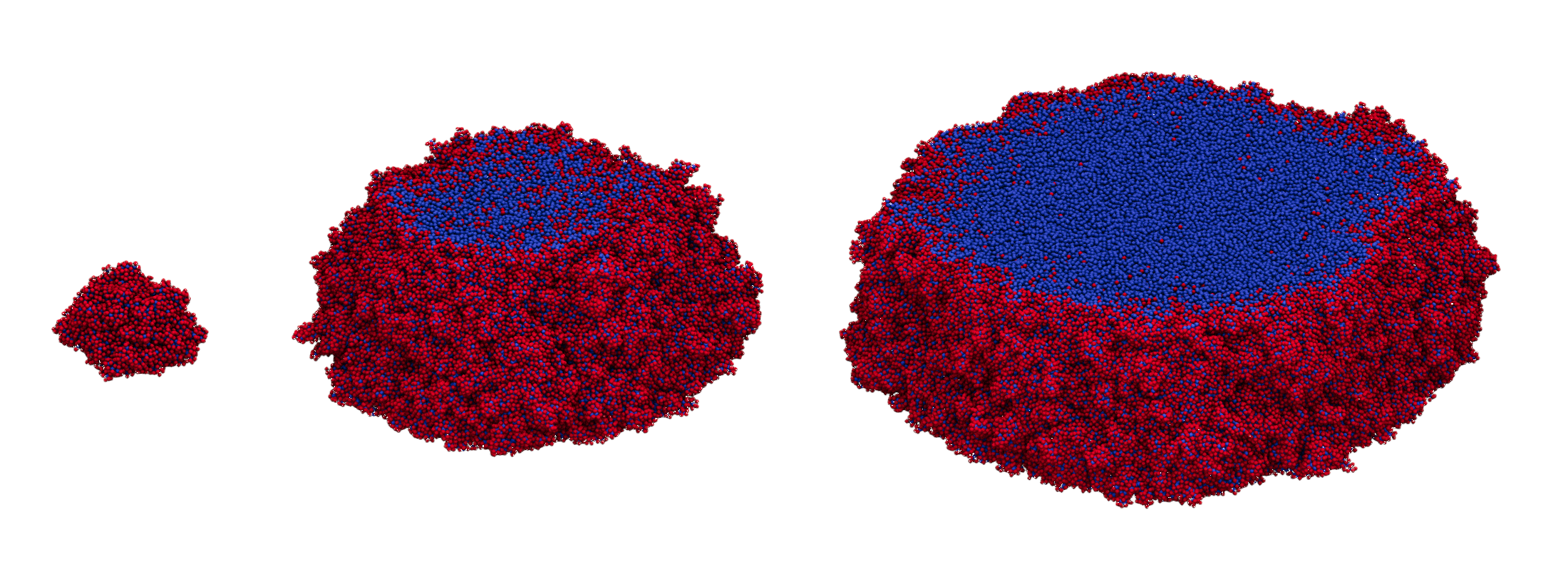}
    \caption{Clusters generated by our model starting from seed cell laid on top of the agar layer.\mmm{The total number of cells in each cluster reads (from left to right): N=3894, 182845, 423239.}}
    \label{fig:modelBTop}
\end{figure}\\
\sk{As shown in Fig.~\ref{fig:modelBTop}, the shape of the cluster growing on the surface of the agar layer is different from that growing from the center only in intermediate states. The change in shape does not translate into a change in quantitative results for the final state, and thus, we performed all further measurements for the original model \textbf{B} growing from the middle of the agar layer.}

The thickness of the cluster is limited by physically imposing boundaries on the growth process.
A thickness constraint expressed in terms of the number of layers was imposed, as shown in Fig.~\ref{fig1-petri}. 
\begin{figure}[!h]
\centering
\includegraphics[width=0.68\columnwidth]{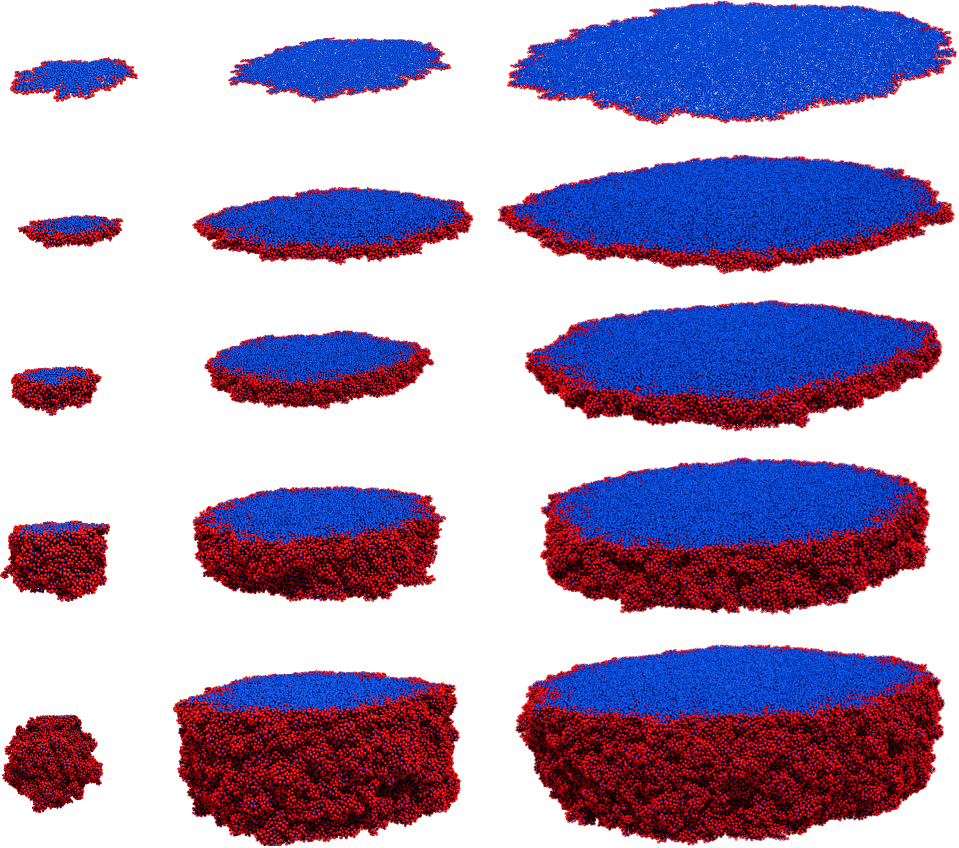}
\caption{Configurations of growing three-dimensional clusters \sk{in time} (from left to right) at a varying thickness of the nutrient layers h=1.6, 5, 10, 20 to 40 (from top to bottom). Red (light in grayscale) cells are alive, whereas blue (dark in grayscale) are dead. \mmm{The total number of cells in each cluster reads N=1343 (left top), 9276, 22704 (right top), 3916, 30931, 80711, 6821, 59567, 160541, 13961, 119685, 324286, 25776 (left bottom), 245830, 657462 (right bottom).}\label{fig3-combo2D}}
\end{figure}
We first simulated several three-dimensional clusters in a Petri dish with an allowed cluster height ranging from h=1 (cluster thickness of one bacterial diameter) to 40. \sk{All simulations continued until the clusters reached the same previously assumed radius equal to 80.} The cluster configurations, represented as the visualization of individual bacteria at different times and thicknesses, are shown in Fig.~\ref{fig3-combo2D}.
Visualization of the cluster configurations at various times and heights revealed their complex and fractal nature, with characteristic cauliflower-like blobs of bacterial colonies on the surface.

\subsubsection{Cluster density}

\begin{figure}[!h]
\centering
\includegraphics[width=0.33\columnwidth]{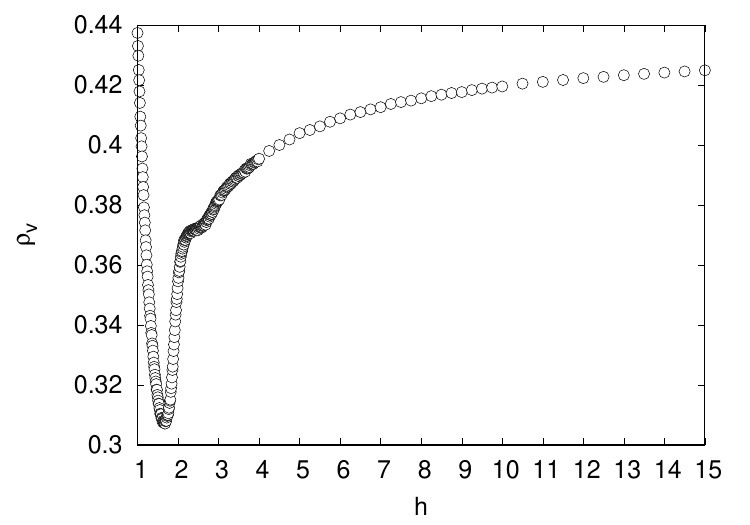}
\includegraphics[width=0.33\columnwidth]{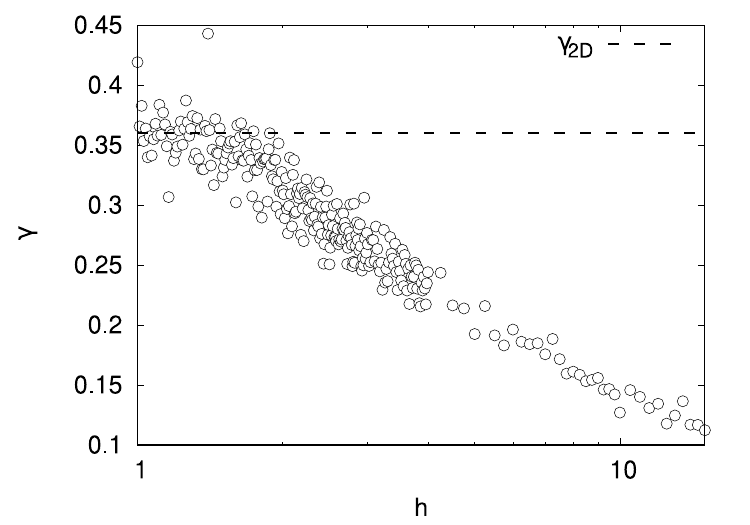}
\includegraphics[width=0.33\columnwidth]{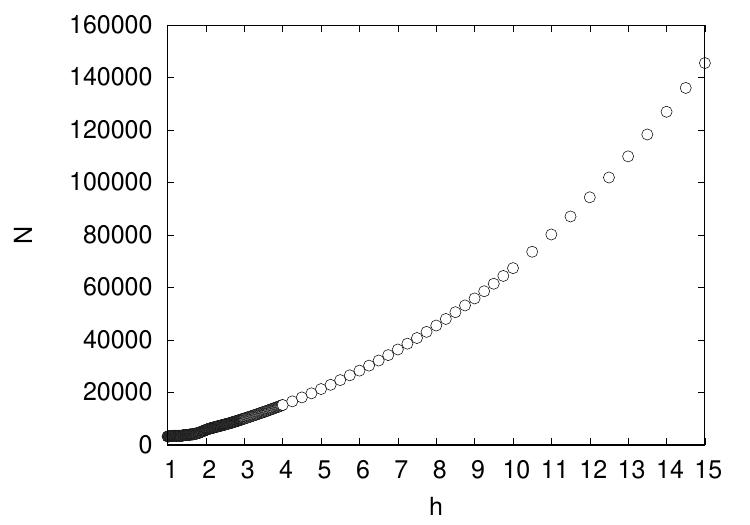}
\caption{Left: Dependence of the density of cells in a cluster on its thickness. The measurement points are the average results from 20 simulations, and the original measurement represents the results for a single cluster. \mmm{Middle:}~$\gamma$ exponent from the formula describing the standard deviation (thickness) of the boundary layer in the cluster $\sigma = \phi R^\gamma$. \mmm{Right: The average number of cells for each height value of simulated clusters. Error bars are smaller than symbols for all three plots.}\label{DENSITY.png}}
\end{figure}

Now, we generalize the way the three-dimensional cluster density is measured. For this, we use the following expression for the volume of the cluster $V_k$: 
\begin{equation} 
    V_k = \pi (Rr)^{2}2rh. 
\end{equation} 
The total volume of the cells $V_c$ is 
\begin{equation} 
    V_c=N \frac{4}{3} \pi r^{3}.
\end{equation} 
The density equals thus: 
\begin{equation}
    \label{cluster_density}
    \rho_v=\frac{V_c}{V_k}=\frac{4}{6} \frac{N}{R^{2}h}.
\end{equation} 
The measurements of the three-dimensional clusters, show that the density $\rho_v$ is constant during the growth process (data not shown) but varies with the cluster thickness (see Fig.~\ref{DENSITY.png}). 
The minima of the density of live cells  near $h \approx 1.6$ indicate that cells pack loosely at this particular height. Additionally, the origin of the kink between $h=2$ and $h=3$ reveals the complex nature of the growth process with a non-trivial combination of at least two various physical mechanisms.

The plot of $\gamma(h)$ (see Fig.~\ref{DENSITY.png} right) shows the behavior of $\gamma$ exponents obtained by fitting the power law $\sigma = \phi R^\gamma$ function from\cite{kardar} to our simulation data. An interesting power-law scaling of the $\gamma$ exponent itself is noticeable. However, we notice that the fitting was most accurate for the lower $h$, and thus, the resulting gamma values for large $h$ should be treated with caution.
\subsection{Fractal dimension of the clusters}

To calculate the fractal dimension of the clusters, we cover the entire space with a cubic mesh of a particular node length. Then, we count the number of boxes overlapping with cluster cells. We repeat this process for several side lengths $s$, each time receiving the number $N(s)$ of boxes. The fractal dimension D can be found using linear regression from the equation:
\begin{equation} 
    \log(N)=-D\log(s)+\log(C),\label{eq:D}
\end{equation}
where $C$ is the free fitting parameter.
For example, for a solid cube object, halving the box size would result in an 8-times increase in the number of filled boxes N. We expect, however, that for clusters, which are similar to flat surfaces with some detailed internal structure, their fractal dimension $D$ will be between two and three. 
 
Our box-count algorithm takes the triplets of bacteria coordinates $(x,y,z)$ and discretizes their shape to fit into the underlying lattice. Then, we set all the lattice elements to be one everywhere, where the bacteria represented as the sphere of radius $r$ overlaps with the lattice. Then, we counted the number of non-zero elements for varying $s$ and used linear regression to Eq.~\ref{eq:D} to find the fractal dimension $D$.
\begin{figure}[!h]
\centering
\includegraphics[width=0.45\columnwidth]{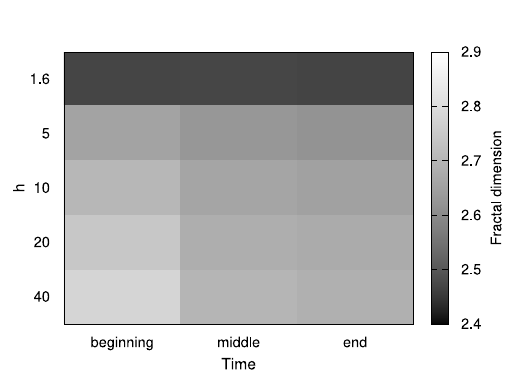}
\caption{The map of fractal dimensions of bacterial colonies measured with the box-counting algorithm.} \label{fractaldim}
\end{figure} 
Looking at the graph (see Fig.~\ref{fractaldim}), we see that the fractal dimension increases with growing thickness and decreases slightly with time of evolution. Our result may be related to the fractal dimension of the Menger Sponge $\log_3{20}\approx2,727$ (see \cite{menger}).

\subsection{Transport properties}

The following subsection treats the bacterial clusters as a porous medium built of tightly packed spherical obstacles. This allows us to study their properties of the transport of gas, fluid, and, in particular, nutrients.

\subsubsection{Permeability}

Permeability is the property of a porous medium that allows fluid to flow through it~\cite{bryant2003permeability}. Permeability depends on the material of the medium, porosity, pore shape, pore connectivity, and many more~\cite{koponen1997permeability, yu2004fractal, shin2015computation}. Moreover, it can vary with conditions imposed on the flow, such as the value of pressure drop~\cite{naaktgeboren2005effect} or temperature~\cite{miadonye2014analytical}. Therefore, many empirical laws relating permeability to other features of porous medium or fluid flow can be found in the literature to make permeability practically useful. Usually, they are based on some assumptions regarding the porous material, the fluid that flows through it, and the flow conditions itself.

We used Ergun's model~\cite{ergun1952fluid} and Kozeny-Carman\cite{carman1956flow} equation to calculate three permeability values for each cluster height. We calculated porosity from the definition of cluster density \eqref{cluster_density} as $1 - \rho_v$. We also used the fact that the cells were modeled as spheres with unit radii. In principle, the model was similar to the classic experimental setup of P. C. Carman ~ \cite{Carman1997fluid}.
 
Ergun's model allows to calculate two values of permeability, 
\begin{equation}
    \label{darcy_permeability}
    K_D = \frac{(2r)^2}{150} \frac{\varepsilon^3}{(1-\varepsilon)^2},
\end{equation}
\begin{equation}
    \label{forvhheimer_permeability}
    K_F = \frac{2r}{1.75} \frac{\varepsilon^3}{1-\varepsilon},
\end{equation}
where $K_D$ and $K_F$ are Darcy's permeability for creeping flow and  Forchheimer's permeability for inertial flow, respectively, $\varepsilon$ is porosity, and $r$ is cell's radius. Kozeny-Carman permeability $K_{KC}$ can is calculated as
\begin{equation}
    \label{kozeny-carman_permeability}
    K_{KC} = \varphi^2 \frac{\varepsilon^3 (2r)^2}{180 (1 - \varepsilon)^2},
\end{equation}
where $\varphi$ is the sphericity of the obstacles that create a porous medium. Because we modeled the cells as spheres of unit radius, $\varphi = 1$.

The permeability results are summarized in Fig.~\ref{fig:permeability_tortusity}. As can be seen, the dependence of permeability $K$ of the cell culture is non-trivial with the height of nutrient-rich layer $h$. In general, the inertial permeability is almost 20 times larger than the Darcy and Kozeny-Carman permeability. Nevertheless, the qualitative characteristics of the permeability dependence on $h$ were the same in all cases. First, when the nutrient-rich layer is not high enough to allow for multiple layers of cells, permeability increases rapidly with increasing $h$. After reaching a maximum for $h=1.68$, the permeability started to decrease at a faster rate than the previously mentioned growth. Next, around $h=2$, a short plateau was created. From $h\approx2.6$, it starts to decrease and converges to a constant value that is approximately $10\%$ higher than the permeability for $h=1$. Our results are in qualitative agreement with a similar experimentally obtained relationship ~\cite{Carman1997fluid}.

\subsubsection{Tortuosity}
Tortuosity $\tau$ can be defined simply as the elongation of fluid particles' pathways as compared with the size of porous medium~\cite{clennell1997tortuosity}
\begin{equation}
    \label{eq:tortuosity_def}
    \tau = \frac{L_a}{L},
\end{equation}
where $L_a$ is the distance that a fluid particle must cover to obtain through the system of length $L$. However, calculating it in real systems, where it depends on the geometry and fluid velocity, may be nontrivial ~\cite{Matyka12}. Therefore, researchers have developed empirical laws that are often used to calculate tortuosity based on the porosity and the shape of obstacles, creating a porous medium.
We calculated the tortuosity from the porosity $\varepsilon$ according to four empirical laws ~\cite{matyka2008tortuosity}.
\begin{equation}
    \label{tortuosity_a}
    \tau_a = \varepsilon^{-p}
\end{equation}
\begin{equation}
    \label{tortuosity_b}
    \tau_b = 1 - p\ln\varepsilon
\end{equation}
\begin{equation}
    \label{tortuosity_c}
    \tau_c = 1 + p(1-\varepsilon)
\end{equation}
\begin{equation}
    \label{tortuosity_d}
    \tau_d = \left(1 + p(1 - \varepsilon)\right)^2
\end{equation}
For equations \eqref{tortuosity_a}, \eqref{tortuosity_b}, \eqref{tortuosity_c}, and \eqref{tortuosity_d}, we use $1$, $1$, $2$, $32/9\pi$ as the values of the parameter $p$, respectively, ~\cite{matyka2008tortuosity}.

\begin{figure}[!h]
\centering
\includegraphics[width=0.45\columnwidth]{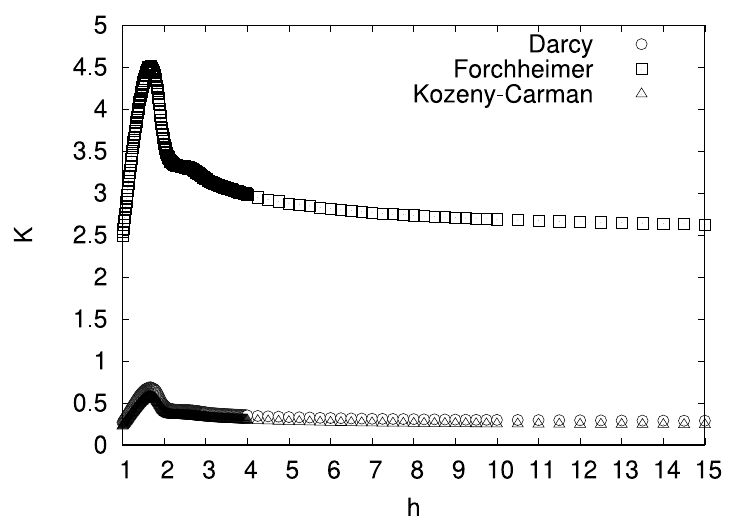}
\includegraphics[width=0.45\columnwidth]{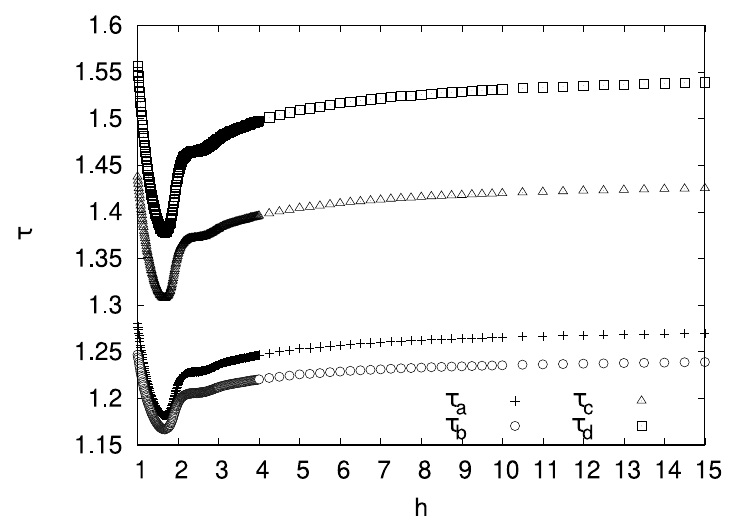}
\caption{Permeability $K$ calculated with three different empirical formulas~\cite{carman1956flow,ergun1952fluid} (left) and Tortuosity $\tau$ calculated with empirical laws mentioned in~\cite{matyka2008tortuosity} (right) versus height of nutrient-rich layer $h$.}
\label{fig:permeability_tortusity}
\end{figure}

The results are shown in Figure~\ref{fig:permeability_tortusity} on the right side. Similar to the permeability results, the dependence of tortuosity on the height of the cell colony can be divided into several regimes. First, we observed a rapid decrease in tortuosity for heights that enabled only single-layer colonies. The minimum tortuosity was determined to be $h=1.68$. Then, the tortuosity increases until $h\approx2$, where it creates a short plateau. For $h>2.6$, the tortuosity increases again and saturates to reach a constant value $1.5\%$ to $2.5\%$ smaller than the initial tortuosity. The overall agreement of tortuosity values to ranges of tortuosities obtained in fluid dynamics and diffusion studies for different porous media is $2.3>\tau>1$ (see\cite{Matyka12, Xu2022}).

\section{Conclusions}

We have developed a new, three-dimensional, off-lattice Eden model to grow thickness-limited clusters. The proposed implementation was first tested with respect to the original two-dimensional implementation. It revealed perfect matching, with only differences arising from the fact that we performed statistical averaging on many clusters. Our implementation consisted of a relatively large number of directions used in the growth process. The three-dimensional model revealed an interesting cauliflower structure of growing clusters, which was previously reported in experiments and confirmed using the continuum model for bacterial colonies\cite{Martinez2022} and revealed in the study of the shape of the Genus Prototheca algae\cite{Rapuntean2009}. We showed the same in the form of pictures of cell configurations in the growth process at various nutrient layer heights. We found that the growth of density with the height of the cluster was nonmonotonous and was the primary driving mechanism for changes in its transport properties, namely, permeability and tortuosity. We analyzed both in terms of empirical laws in the case of slow and inertial transport regimes. Our permeability results agree qualitatively with the experimental setup of packed granular beds~\cite{Carman1997fluid}. We also analyzed the morphology of clusters in terms of fractal dimension and revealed their similarity to the Menger Sponge for particular cluster heights. Our findings may help understand the origin of the morphology and transport properties of experimentally grown bacterial clusters. We also believe that our results may not only be useful in the study of bacterial colonies but also in learning about the properties of cell clusters\cite{Frey20}. However, to use our relationships and conclusions, they should be verified under laboratory conditions.

\section*{Acknowledgements}

We are grateful to Dawid Strzelczyk for discussions at an early stage of this work and to prof. Zbigniew Koza for valuable comments.
Funded by National Science Centre, Poland under the OPUS call in the Weave programme 2021/43/I/ST3/00228.
This research was funded in whole or in part by National Science Centre (2021/43/I/ST3/00228). For the purpose of Open Access,
the author has applied a CC-BY public copyright licence to any Author Accepted Manuscript (AAM) version arising from this submission.
\mm{M. Matyka acknowledges the financial support from the Slovenian Research And Innovation Agency (ARIS) research core funding No.\ P2-0095.}
\mmm{Authors would like to thank an anonymous reviewer for constructive evaluation of our work that helped us to improve the manuscript and results presented.}


\begin{thebibliography}{10}
\urlstyle{rm}
\expandafter\ifx\csname url\endcsname\relax
  \def\url#1{\texttt{#1}}\fi
\expandafter\ifx\csname urlprefix\endcsname\relax\def\urlprefix{URL }\fi
\expandafter\ifx\csname doiprefix\endcsname\relax\def\doiprefix{DOI: }\fi
\providecommand{\bibinfo}[2]{#2}
\providecommand{\eprint}[2][]{\url{#2}}

\bibitem{bacteriafood}
\bibinfo{author}{De~Vuyst, L.} \& \bibinfo{author}{Leroy, F.}
\newblock \bibinfo{journal}{\bibinfo{title}{{Bacteriocins from Lactic Acid Bacteria: Production, Purification, and Food Applications}}}.
\newblock {\emph{\JournalTitle{Journal of Molecular Microbiology and Biotechnology}}} \textbf{\bibinfo{volume}{13}}, \bibinfo{pages}{194--199}, \doiprefix\url{10.1159/000104752} (\bibinfo{year}{2007}).
\newblock \eprint{https://karger.com/mmb/article-pdf/13/4/194/4138561/000104752.pdf}.

\bibitem{aneja2007experiments}
\bibinfo{author}{Aneja, K.}
\newblock \emph{\bibinfo{title}{Experiments In Microbiology, Plant Pathology And Biotechnology}} (\bibinfo{publisher}{New Age International}, \bibinfo{year}{2007}).

\bibitem{Chiang2020}
\bibinfo{author}{Chiang, T.-C.}, \bibinfo{author}{Huang, M.-S.}, \bibinfo{author}{Lu, P.-L.}, \bibinfo{author}{Huang, S.-T.} \& \bibinfo{author}{Lin, Y.-C.}
\newblock \bibinfo{journal}{\bibinfo{title}{The effect of oral care intervention on pneumonia hospitalization, staphylococcus aureus distribution, and salivary bacterial concentration in taiwan nursing home residents: a pilot study}}.
\newblock {\emph{\JournalTitle{BMC Infectious Diseases}}} \textbf{\bibinfo{volume}{20}}, \doiprefix\url{10.1186/s12879-020-05061-z} (\bibinfo{year}{2020}).

\bibitem{Singh2017}
\bibinfo{author}{Singh, M.}, \bibinfo{author}{Kumar, A.}, \bibinfo{author}{Singh, R.} \& \bibinfo{author}{Pandey, K.~D.}
\newblock \bibinfo{journal}{\bibinfo{title}{Endophytic bacteria: a new source of bioactive compounds}}.
\newblock {\emph{\JournalTitle{3 Biotech}}} \textbf{\bibinfo{volume}{7}}, \bibinfo{pages}{1--14} (\bibinfo{year}{2017}).

\bibitem{Medema2011}
\bibinfo{author}{Medema, M.~H.}, \bibinfo{author}{Breitling, R.} \& \bibinfo{author}{Takano, E.}
\newblock \bibinfo{title}{Synthetic biology in streptomyces bacteria}.
\newblock In \emph{\bibinfo{booktitle}{Methods in enzymology}}, vol. \bibinfo{volume}{497}, \bibinfo{pages}{485--502} (\bibinfo{publisher}{Elsevier}, \bibinfo{year}{2011}).

\bibitem{Shapiro95}
\bibinfo{author}{Shapiro, J.~A.}
\newblock \bibinfo{journal}{\bibinfo{title}{The significances of bacterial colony patterns}}.
\newblock {\emph{\JournalTitle{BioEssays}}} \textbf{\bibinfo{volume}{17}}, \bibinfo{pages}{597--607} (\bibinfo{year}{1995}).

\bibitem{Wimpenny79}
\bibinfo{author}{Wimpenny, J.~W.}
\newblock \bibinfo{journal}{\bibinfo{title}{The growth and form of bacterial colonies}}.
\newblock {\emph{\JournalTitle{Microbiology}}} \textbf{\bibinfo{volume}{114}}, \bibinfo{pages}{483--486} (\bibinfo{year}{1979}).

\bibitem{Ben1994}
\bibinfo{author}{Ben-Jacob, E.} \emph{et~al.}
\newblock \bibinfo{journal}{\bibinfo{title}{Generic modelling of cooperative growth patterns in bacterial colonies}}.
\newblock {\emph{\JournalTitle{Nature}}} \textbf{\bibinfo{volume}{368}}, \bibinfo{pages}{46--49} (\bibinfo{year}{1994}).

\bibitem{Li1995}
\bibinfo{author}{Li, B.}, \bibinfo{author}{Wang, J.}, \bibinfo{author}{Wang, B.}, \bibinfo{author}{Liu, W.} \& \bibinfo{author}{Wu, Z.}
\newblock \bibinfo{journal}{\bibinfo{title}{Computer simulations of bacterial-colony formation}}.
\newblock {\emph{\JournalTitle{Europhysics Letters}}} \textbf{\bibinfo{volume}{30}}, \bibinfo{pages}{239} (\bibinfo{year}{1995}).

\bibitem{Grimson}
\bibinfo{author}{Grimson, M.~J.} \& \bibinfo{author}{Barker, G.~C.}
\newblock \bibinfo{journal}{\bibinfo{title}{A continuum model for the growth of bacterial colonies on a surface}}.
\newblock {\emph{\JournalTitle{Journal of Physics A: Mathematical and General}}} \textbf{\bibinfo{volume}{26}}, \bibinfo{pages}{5645} (\bibinfo{year}{1993}).

\bibitem{Eden61}
\bibinfo{author}{Eden, M.}
\newblock \bibinfo{journal}{\bibinfo{title}{A two-dimensional growth process}}.
\newblock {\emph{\JournalTitle{Dynamics of fractal surfaces}}} \textbf{\bibinfo{volume}{4}}, \bibinfo{pages}{598} (\bibinfo{year}{1961}).

\bibitem{Witten81}
\bibinfo{author}{Witten, T.~A.} \& \bibinfo{author}{Sander, L.~M.}
\newblock \bibinfo{journal}{\bibinfo{title}{Diffusion-limited aggregation, a kinetic critical phenomenon}}.
\newblock {\emph{\JournalTitle{Phys. Rev. Lett.}}} \textbf{\bibinfo{volume}{47}}, \bibinfo{pages}{1400--1403}, \doiprefix\url{10.1103/PhysRevLett.47.1400} (\bibinfo{year}{1981}).

\bibitem{AUSLOOS19952185}
\bibinfo{author}{Ausloos, M.}, \bibinfo{author}{Vandewalle, N.} \& \bibinfo{author}{Cloots, R.}
\newblock \bibinfo{journal}{\bibinfo{title}{Magnetic kinetic growth models}}.
\newblock {\emph{\JournalTitle{Journal of Magnetism and Magnetic Materials}}} \textbf{\bibinfo{volume}{140-144}}, \bibinfo{pages}{2185--2186}, \doiprefix\url{https://doi.org/10.1016/0304-8853(94)00542-7} (\bibinfo{year}{1995}).
\newblock \bibinfo{note}{International Conference on Magnetism}.

\bibitem{Wang1995}
\bibinfo{author}{Wang, C.~Y.}, \bibinfo{author}{Liu, P.-L.} \& \bibinfo{author}{Bassingthwaighte, J.}
\newblock \bibinfo{journal}{\bibinfo{title}{Off-lattice eden-c cluster growth model}}.
\newblock {\emph{\JournalTitle{Journal of physics A: Mathematical and general}}} \textbf{\bibinfo{volume}{28}}, \bibinfo{pages}{2141} (\bibinfo{year}{1995}).

\bibitem{monod}
\bibinfo{author}{Monod, J.}
\newblock \bibinfo{journal}{\bibinfo{title}{The growth of bacterial cultures}}.
\newblock {\emph{\JournalTitle{Annual Review of Microbiology}}} \textbf{\bibinfo{volume}{3}}, \bibinfo{pages}{371--394}, \doiprefix\url{https://doi.org/10.1146/annurev.mi.03.100149.002103} (\bibinfo{year}{1949}).

\bibitem{code}
\bibinfo{author}{Kaczmarczyk, S.}
\newblock \bibinfo{title}{{PetriDishModel: Off-lattice Eden-C Simulation Model for bacteria growth in a Petri Dish}}.
\newblock \bibinfo{howpublished}{\url{https://github.com/Kamikadzerek/PetriDishModel}} (\bibinfo{year}{2024}).
\newblock \bibinfo{note}{Accessed: 2024-05-31}.

\bibitem{kardar}
\bibinfo{author}{Kardar, M.}, \bibinfo{author}{Parisi, G.} \& \bibinfo{author}{Zhang, Y.-C.}
\newblock \bibinfo{journal}{\bibinfo{title}{Dynamic scaling of growing interfaces}}.
\newblock {\emph{\JournalTitle{Phys. Rev. Lett.}}} \textbf{\bibinfo{volume}{56}}, \bibinfo{pages}{889--892}, \doiprefix\url{10.1103/PhysRevLett.56.889} (\bibinfo{year}{1986}).

\bibitem{Martinez2022}
\bibinfo{author}{Mart{\'\i}nez-Calvo, A.} \emph{et~al.}
\newblock \bibinfo{journal}{\bibinfo{title}{Morphological instability and roughening of growing 3d bacterial colonies}}.
\newblock {\emph{\JournalTitle{Proceedings of the National Academy of Sciences}}} \textbf{\bibinfo{volume}{119}}, \bibinfo{pages}{e2208019119} (\bibinfo{year}{2022}).

\bibitem{Huergo2012}
\bibinfo{author}{Huergo, M. A.~C.}, \bibinfo{author}{Pasquale, M.~A.}, \bibinfo{author}{Gonz\'alez, P.~H.}, \bibinfo{author}{Bolz\'an, A.~E.} \& \bibinfo{author}{Arvia, A.~J.}
\newblock \bibinfo{journal}{\bibinfo{title}{Growth dynamics of cancer cell colonies and their comparison with noncancerous cells}}.
\newblock {\emph{\JournalTitle{Phys. Rev. E}}} \textbf{\bibinfo{volume}{85}}, \bibinfo{pages}{011918}, \doiprefix\url{10.1103/PhysRevE.85.011918} (\bibinfo{year}{2012}).

\bibitem{Colter2001}
\bibinfo{author}{Colter, D.~C.}, \bibinfo{author}{Sekiya, I.} \& \bibinfo{author}{Prockop, D.~J.}
\newblock \bibinfo{journal}{\bibinfo{title}{Identification of a subpopulation of rapidly self-renewing and multipotential adult stem cells in colonies of human marrow stromal cells}}.
\newblock {\emph{\JournalTitle{Proceedings of the National Academy of Sciences}}} \textbf{\bibinfo{volume}{98}}, \bibinfo{pages}{7841--7845}, \doiprefix\url{10.1073/pnas.141221698} (\bibinfo{year}{2001}).
\newblock \eprint{https://www.pnas.org/doi/pdf/10.1073/pnas.141221698}.

\bibitem{Kaufman2001}
\bibinfo{author}{Kaufman, D.~S.}, \bibinfo{author}{Hanson, E.~T.}, \bibinfo{author}{Lewis, R.~L.}, \bibinfo{author}{Auerbach, R.} \& \bibinfo{author}{Thomson, J.~A.}
\newblock \bibinfo{journal}{\bibinfo{title}{Hematopoietic colony-forming cells derived from human embryonic stem cells}}.
\newblock {\emph{\JournalTitle{Proceedings of the National Academy of Sciences}}} \textbf{\bibinfo{volume}{98}}, \bibinfo{pages}{10716--10721}, \doiprefix\url{10.1073/pnas.191362598} (\bibinfo{year}{2001}).
\newblock \eprint{https://www.pnas.org/doi/pdf/10.1073/pnas.191362598}.

\bibitem{menger}
\bibinfo{author}{Bunde, A.} \& \bibinfo{author}{Havlin, S.}
\newblock \emph{\bibinfo{title}{Fractals in Science}}.
\newblock Fractals in Science (\bibinfo{publisher}{Springer Berlin Heidelberg}, \bibinfo{year}{2013}).

\bibitem{bryant2003permeability}
\bibinfo{author}{Bryant, W.~R.}
\newblock \emph{\bibinfo{title}{Permeability of Clays, Silty-Clays and Clayey-Silts}}, \bibinfo{pages}{344–403} (\bibinfo{publisher}{SEPM Society for Sedimentary Geology}).

\bibitem{koponen1997permeability}
\bibinfo{author}{Koponen, A.}, \bibinfo{author}{Kataja, M.} \& \bibinfo{author}{Timonen, J.}
\newblock \bibinfo{journal}{\bibinfo{title}{Permeability and effective porosity of porous media}}.
\newblock {\emph{\JournalTitle{Physical Review E}}} \textbf{\bibinfo{volume}{56}}, \bibinfo{pages}{3319} (\bibinfo{year}{1997}).

\bibitem{yu2004fractal}
\bibinfo{author}{Yu, B.} \& \bibinfo{author}{Liu, W.}
\newblock \bibinfo{journal}{\bibinfo{title}{Fractal analysis of permeabilities for porous media}}.
\newblock {\emph{\JournalTitle{AIChE journal}}} \textbf{\bibinfo{volume}{50}}, \bibinfo{pages}{46--57} (\bibinfo{year}{2004}).

\bibitem{shin2015computation}
\bibinfo{author}{Shin, H.~S.}, \bibinfo{author}{Kim, K.~Y.} \& \bibinfo{author}{Pande, G.~N.}
\newblock \bibinfo{journal}{\bibinfo{title}{On computation of strain-dependent permeability of rocks and rock-like porous media}}.
\newblock {\emph{\JournalTitle{International Journal for Numerical and Analytical Methods in Geomechanics}}} \textbf{\bibinfo{volume}{39}}, \bibinfo{pages}{821--832} (\bibinfo{year}{2015}).

\bibitem{naaktgeboren2005effect}
\bibinfo{author}{Naaktgeboren, C.}, \bibinfo{author}{Krueger, P.~S.} \& \bibinfo{author}{Lage, J.~L.}
\newblock \bibinfo{title}{The effect of inlet and exit pressure-drop on the determination of porous media permeability and form coefficient}.
\newblock In \emph{\bibinfo{booktitle}{Fluids Engineering Division Summer Meeting}}, vol. \bibinfo{volume}{41987}, \bibinfo{pages}{205--211} (\bibinfo{year}{2005}).

\bibitem{miadonye2014analytical}
\bibinfo{author}{Miadonye, A.} \& \bibinfo{author}{Amadu, M.}
\newblock \bibinfo{journal}{\bibinfo{title}{Analytical derivation of the temperature dependence of absolute permeability of a porous medium}}.
\newblock {\emph{\JournalTitle{Journal of Petroleum Science Research (JPSR) Volume}}} \textbf{\bibinfo{volume}{3}} (\bibinfo{year}{2014}).

\bibitem{ergun1952fluid}
\bibinfo{author}{Ergun, S.}
\newblock \bibinfo{journal}{\bibinfo{title}{Fluid flow through packed columns}}.
\newblock {\emph{\JournalTitle{Chemical engineering progress}}} \textbf{\bibinfo{volume}{48}}, \bibinfo{pages}{89} (\bibinfo{year}{1952}).

\bibitem{carman1956flow}
\bibinfo{author}{Carman, P.}
\newblock \emph{\bibinfo{title}{Flow of Gases Through Porous Media}} (\bibinfo{publisher}{Academic Press}, \bibinfo{year}{1956}).

\bibitem{Carman1997fluid}
\bibinfo{author}{Carman, P.~C.}
\newblock \bibinfo{journal}{\bibinfo{title}{Fluid flow through granular beds}}.
\newblock {\emph{\JournalTitle{Chemical Engineering Research and Design}}} \textbf{\bibinfo{volume}{75}}, \bibinfo{pages}{S32--S48} (\bibinfo{year}{1997}).

\bibitem{clennell1997tortuosity}
\bibinfo{author}{Clennell, M.~B.}
\newblock \bibinfo{journal}{\bibinfo{title}{Tortuosity: a guide through the maze}}.
\newblock {\emph{\JournalTitle{Geological Society, London, Special Publications}}} \textbf{\bibinfo{volume}{122}}, \bibinfo{pages}{299--344} (\bibinfo{year}{1997}).

\bibitem{Matyka12}
\bibinfo{author}{Matyka, M.} \& \bibinfo{author}{Koza, Z.}
\newblock \bibinfo{title}{How to calculate tortuosity easily?}
\newblock In \emph{\bibinfo{booktitle}{AIP Conference Proceedings 4}}, vol. \bibinfo{volume}{1453}, \bibinfo{pages}{17--22} (\bibinfo{organization}{American Institute of Physics}, \bibinfo{year}{2012}).

\bibitem{matyka2008tortuosity}
\bibinfo{author}{Matyka, M.}, \bibinfo{author}{Khalili, A.} \& \bibinfo{author}{Koza, Z.}
\newblock \bibinfo{journal}{\bibinfo{title}{Tortuosity-porosity relation in porous media flow}}.
\newblock {\emph{\JournalTitle{Physical Review E}}} \textbf{\bibinfo{volume}{78}}, \bibinfo{pages}{026306} (\bibinfo{year}{2008}).

\bibitem{Xu2022}
\bibinfo{author}{Xu, W.}, \bibinfo{author}{Zhang, K.}, \bibinfo{author}{Zhang, Y.} \& \bibinfo{author}{Jiang, J.}
\newblock \bibinfo{journal}{\bibinfo{title}{Packing fraction, tortuosity, and permeability of granular-porous media with densely packed spheroidal particles: Monodisperse and polydisperse systems}}.
\newblock {\emph{\JournalTitle{Water Resources Research}}} \textbf{\bibinfo{volume}{58}}, \bibinfo{pages}{e2021WR031433} (\bibinfo{year}{2022}).

\bibitem{Rapuntean2009}
\bibinfo{author}{Rapuntean, S.}, \bibinfo{author}{Rapuntean, G.}, \bibinfo{author}{FIT, N.~I.}, \bibinfo{author}{Cosmina, C.} \& \bibinfo{author}{NADAS, G.~C.}
\newblock \bibinfo{journal}{\bibinfo{title}{Morphological and cultural characterization of some strains of unicellular algae of the genus prototheca sampled from mastitic cow milk}}.
\newblock {\emph{\JournalTitle{Notulae Botanicae Horti Agrobotanici Cluj-Napoca}}} \textbf{\bibinfo{volume}{37}}, \bibinfo{pages}{31--40} (\bibinfo{year}{2009}).

\bibitem{Frey20}
\bibinfo{author}{Frey, F.} \emph{et~al.}
\newblock \bibinfo{journal}{\bibinfo{title}{Eden growth models for flat clathrin lattices with vacancies}}.
\newblock {\emph{\JournalTitle{New Journal of Physics}}} \textbf{\bibinfo{volume}{22}}, \bibinfo{pages}{073043} (\bibinfo{year}{2020}).

\end{thebibliography}
\bibliographystyle{elsarticle-num}

\end{document}